# Monte Carlo simulation and parameterized treatment on the effect of nuclear elastic scattering in high-energy proton radiography[*]

XU Hai-Bo（许海波）　　ZHENG Na（郑娜）[1)]

Institute of Applied Physics and Computational Mathematics, Beijing 100094, China

**Abstract:** A version of Geant4 has been developed to treat high-energy proton radiography. This article presents the results of calculations simulating the effects of nuclear elastic scattering for various test step wedges. Comparisons with experimental data are also presented. The traditional expressions of the transmission should be correct if the angle distribution of the scattering is Gaussian multiple Coulomb scattering. The mean free path which depends on the collimator angle and the radiation length are treated as empirical parameters, according to transmission as a function of thickness obtained by simulations. The results benefit for reconstructing density that depends on the transmission expressions.

**Key words:** proton radiography, multiple Coulomb scattering, Angle collimator, nuclear elastic scattering, Geant4

**PACS:** 29.27.Eg

## 1　Introduction

　　High-energy proton radiography offers a promising new diagnostic technique to determine the geometric configuration and physical characteristic related to primary implosion phenomena because the mean free path can be tailored to allow seeing inside almost any experiment [1]. Proton radiography has been shown that is far superior to flash x-ray radiography. In proton radiography, the attenuation of protons passing through an object is measured by focusing the transmitted beam with a set of quadrupole magnets onto an image plane detector. Protons interact with matter by way of strong and electromagnetic interactions. The three most important effects on the protons as they go through an object are absorption, multiple Coulomb scattering, and energy loss [2-4]. First, some of the protons are absorbed by nuclear collisions. This is a simple exponential attenuation of the beam. Second, the protons are scattered into small angles by multiple Coulomb scattering. This not only produces image blurring, but it also changes the total attenuation as a detector may not see the full phase space of the scattered beam. Third, protons lose varying amounts of energy as they go through an object from both energy straggling and thickness variations in the object. This produces a spread in the momentum of the transmitted protons that blurs the final image due to chromatic aberrations in the lens.

---

* Supported by NSAF (11176001) and Science and Technology Developing Foundation of China Academy of Engineering Physics (2012A0202006)
1) E-mail：nnazheng@gmail.com.

For full-scale hydrotest experiments or thick objects, the angular spread of the beam fraction that passes through the magnetic lens to form the image is dominated by multiple Coulomb scattering (MCS) in the radiographed object, but elastic proton–nucleon scattering in the object can also have a significant effect on the transmitted angular distribution [5,6].

The traditional expressions of the transmission should be correct if the angle distribution of the scattering is Gaussian MCS. In determining volume density distributions by fitting parametric models of the objects to experimental radiographs, the mean free path which depends on the collimator angle and the radiation length are treated as empirical parameters. In this paper, the net attenuation of the illuminating beamlets by angle-cut collimation will be emphasized.

## 2  The basic equation of pronton radiography

A simple model for proton radiography can been obtained by assuming that nuclear attenuation in the removal of beam particles from the transmitted flux and that Coulomb scattering can approximated by assuming transmitted particles are scattered into a Gaussian shaped angular distribution [1-4,7-9]. In this approximation integration of the angular distribution between limits imposed by angle collimators at the Fourier points in the lenses results in closed form expressions for the transmission. This gives:

$$T(L) = \exp\left(-\sum_i \frac{L_i}{\lambda_i}\right)\left[1 - \exp\left(-\frac{\theta_{\text{cut}}^2}{2\theta_0^2}\right)\right] = \exp\left(-\sum_i \frac{L_i}{\lambda_i}\right)\left\{1 - \exp\left[-\kappa\Big/\left(\sum_i L_i/X_{0i}\right)\right]\right\} \quad (1)$$

Here, $L_i$ is the areal density for the $i'th$ material, $\lambda_i$ is the nuclear attenuation factor for the $i'th$ material given by

$$\lambda_i = \frac{A}{N_A \sigma_i} \quad (2)$$

where $N_A$ is is Avogadro's number, $A_i$ is the atomic weight and $\sigma_i$ is the absorption cross section for the *i'th* material.

And $\theta_{cut}$ is the magnitude cut imposed by the angle-cut collimator and $\theta_0$ is the multiple coulomb scattering angle given approximately by

$$\theta_0 \approx \frac{14.1\,\text{MeV}}{pc\beta}\sqrt{\sum_i^n \frac{L_i}{X_{0i}}} \tag{3}$$

Here, $p$ is the beam momentum, $\beta = v/c$ where $v$ is the beam velocity and $c$ is the speed of light, $X_{0i}$ is the radiation length for the *i'th* material given by

$$X_{0i} = \frac{716.4 A_i}{Z_i(Z_i+1)\ln(287/\sqrt{Z_i})} \tag{4}$$

The scale $\kappa$,

$$\kappa = \frac{1}{2}\left(\frac{pc\beta\theta_{cut}}{14.1\,\text{MeV}}\right)^2 \tag{5}$$

is independent of *L* and the material.

The effect of beam emittance on the transmission can be included in this Gaussian approximation by adding the additional contribution to the beam divergence referenced to the offset *a*, normalized to units of $\kappa$, in quadrature to the multiple scattering caused by the object[6,7]. This gives

$$T(L) = \exp\left(-\sum_i \frac{L_i}{\lambda_i}\right)\left[1-\exp\left(-\frac{\theta_{cut}^2}{2\theta_0^2}\right)\right] = \exp\left(-\sum_i \frac{L_i}{\lambda_i}\right)\left\{1-\exp\left[-\kappa\bigg/\left(\sum_i L_i/X_{0i}+a\right)\right]\right\} \tag{6}$$

The relationship between the divergence of the beam $\theta_b$, and the parameter *a* is

$$a = \left(\frac{pc\beta\theta_b}{14.1\,\text{MeV}}\right)^2 \tag{7}$$

## 3  The simulation of step wedge based on Geant4 toolkit

### 3.1  The setup of simulation

To test the proton radiography concept at higher energies, the experiment E955 for the beam of 24 GeV/c protons provided by the Alternating Gradient Synchrotron (AGS) at Brookhaven National Laboratory involved a series of transmission measurements as a function of material thickness and position on various objects. According to the experiment, we made serious simulations, and the components of the simulation setup are shown in Fig. 1. The beam is first prepared with a diffuser and matching lens to meet optics requirements. Next the beam is measured just upstream of the object by the front detectors after which it passes through the object being radiographed. The beam extracted from the AGS was focused onto a 1.2 cm-thick tantalum diffuser. The lens consisted of 4 quadrupoles, 20 cm diameter, 120 cm long, that were configured to form a unit magnification imaging lens. Either of a pair of collimators, 1.2 m long right circular cylinders of tungsten, was located at the Fourier mid-plane of the lens. The collimators approximated multiple-scattering angle acceptance cuts of 6.68 mrad [6]. In order to do comparisons with experimental data, a set of step wedges was radiographed in the simulation with 24 GeV/c and $10^{10}$ incident protons.

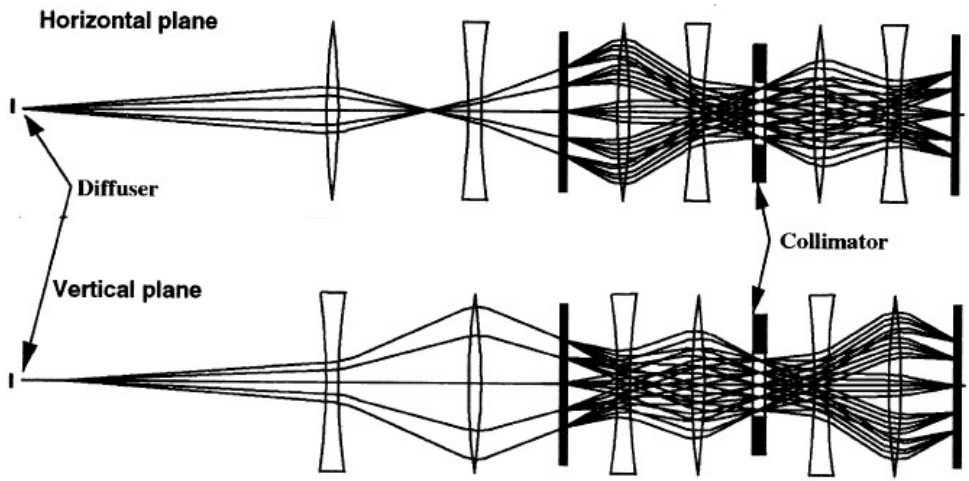

Fig. 1. Layout of the proton radiography magnetic lens system used for simulation.

**3.2  The transmission of the step wedge**

Numerical simulation using the Monte Carlo code Geant4 [10] has been implemented to investigate the entire physics mechanism. In the first step, the proper functioning of the Geant4 code is checked by comparing simulation results with measurement results. The calculated results and the experimental results are shown in Figs. 3 and 4. The calculated results and the experimental results are found to be in good agreement.

The central part of the collimator description in the simulation was the same as the experimental set up. Rather than put the object at the object location in Fig.1, there was a mesh tally followed by a –I lens (lens 0), the object was centered at the image plane of lens 0, but this collimator did not intercept any protons [11,12].

These value of transmission versus areal density are plotted in Figs. 2 and 3 for carbon and lead, respectively, where the experimental measurements [13] are also plotted. It was pointed out that inclusion of nuclear scattering with the Gaussian MCS would improve the agreement between

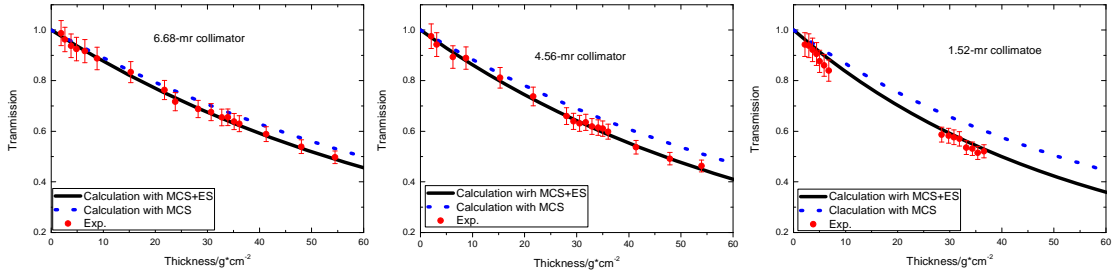

Monte Carlo simulations and the data.

Fig. 2. Transmissions as a function of thickness for carbon compared with the experiment data for three different

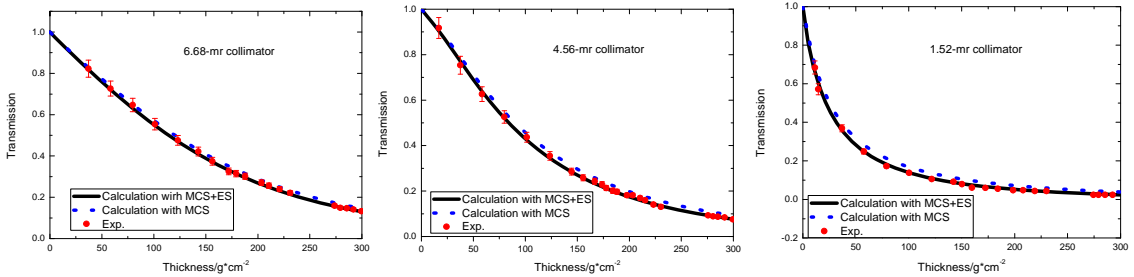

angle-cut collimators.

Fig. 3. Transmissions as a function of thickness for lead compared with the experiment data for three different angle-cut collimators

## 4  The parameterized treatment on the effect of nuclear elastic scattering

For thick objects, the angular spread of the beam fraction that passes through the magnetic lens to form the image is dominated by MCS in the radiographed object, but elastic proton–nucleon scattering in the object can also have a significant effect on the transmitted angular distribution. The above expressions should be corrected if the angular distribution of the scattering is Gaussian MCS. The step wedge data the experiment could be not fit for the three collimators and a given material, with the same values for $\lambda$ and $X_0$. In order to fit the data, Morris C L et al. proposed a method that for different collimators with one value of $X_0$, an angle-cut dependent value of the nuclear interaction length $\lambda$ was assumed [6]. This result, while possibly functional, was not pleasing because of the addition of new parameters.

Proton radiography requires a high energy beam to penetrate a thick object while keeping the MCS angle and energy loss small enough to allow good spatial resolution. In order to satisfy the goal that the quantitative radiography of a dynamic experiment with plutonium, the accelerator design requirements is 50 GeV/c protons for the Advanced Hydrotest Facility (AHF) by United States. Proton momentum of 50 GeV/c is suitable to the design goal of 1 mm imaging blur. On the other hand, in order to do comparisons with experimental data provided by the AGS at BNL. The nominal proton momentum in E955 was 24 GeV/c.

In this paper, we introduced a scenario of 24 GeV/c and 50 GeV/c proton radiography beamline, in which a matching section, the Zumbro lens (minus identity) lens system and imaging system will be particularized. Parameters of the minus identity lens for 24GeV/c and 50 GeV/c proton radiography are given in Table 1.

Table 1. Parameters of the minus identity lens for 24GeV/c and 50 GeV/c proton radiography.

| momentum /GeV·c$^{-1}$ | quadrupole aperture/mm | quadrupole gradient /T·m$^{-1}$ | quadrupole length/m | drift length/m | chromatic aberration coefficient/m | field of view/mm |
|---|---|---|---|---|---|---|
| 24 | 120 | 8 | 2 | 3.4 | 40.4 | 80 |
| 50 | 220 | 16.74 | 3 | 4 | 36.1 | 120 |

Transmission as a function of thickness was obtained by Monte Carlo simulations based on Geant4. Step wedges were imaged to obtain transmission as a function of thickness for eight materials: beryllium, carbon, aluminum, iron, copper, tin, tungsten, and lead. These materials were chosen so that our parameterization would span the periodic table. Results for 24 GeV/c and 50 GeV/c protons are shown in Figs. 4 and 5.

The lines show least-squares fits to the data using Eq. (6). If $\lambda_\theta$ and $X_0$ are treated as empirical parameters-good fits to transmissions-a function of thickness can be obtained.

In the fit, the $X_0$ were parameterized by

$$X_0 = \frac{a_x A}{Z(Z+1)\ln\left(b_x/\sqrt{Z}\right)} \tag{8}$$

The $\lambda_\theta$ were parameterized as

$$\lambda_\theta = a_\theta A^{b_\theta} \tag{9}$$

for collimator angles $\theta$. Here $A$ and $Z$ are the atomic weight and the atomic number, respectively. The material dependence is entirely contained in the $A$ and $Z$ in Eqs. (8) and (9).

In this fit, the independent parameters are $L_{i,j}$, the areal density of of each step $j$ of $i$'th material. The dependent parameters are the measured transmissions $T_{i,j}$ through each step. The material properties $A_i$ and $Z_i$, the beam momentum $p\beta$, and the collimator cut angles $\theta_{cut}$ are taken as known constants. This leaves seven parameters to fit: the four collimator-dependent cross section parameters $a_\theta$ and $b_\theta$ from Eq. (9); the two radiation length parameters $a_X$ and $b_X$ from Eq. (8); and the beam divergence $\theta_b$ from Eq. (7).

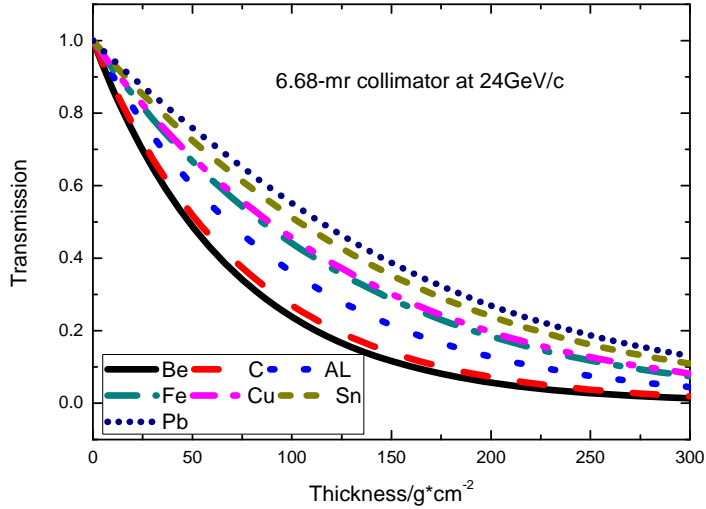

FIG. 4. The calculated transmissions as a function of thickness for the step wedges data at 24GeV/c.

This function can be seen to give a good account of the data over a very wide range in step wedge thickness. In the step wedge fits, below, we will parameterize $X_0$, and fit for the effective beam emittance $\theta_b$.

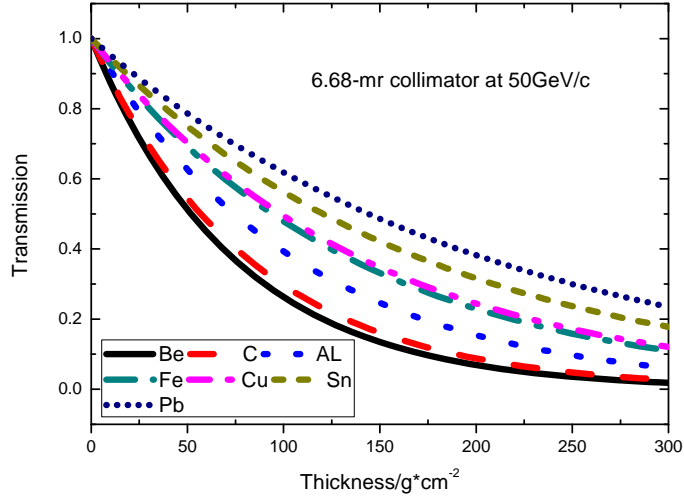

Fig. 5. The calculated transmissions as a function of thickness for the step wedge data at 50GeV/c.

For the 24GeV/c and 50 GeV/c beamline, the results of the fit are given in Table 2. From table 2, we can see that our results are good for the agreement with that of the ref. [6].

Table 2. Fitted parameters for the radiation length and attenuation length parameterization for 24GeV/c and 50 GeV/c.

| Momentum(GeV/c) | | $a_\theta$ (g/cm$^2$) | $b_\theta$ | $a_X$ (g/cm$^2$) | $b_X$ | $\theta_b$ (mr) |
|---|---|---|---|---|---|---|
| 24 | results of [6] | 35.06 | 0.3182 | 230.1 | 34.5 | 1.05 |
|  | our result | 34.67 | 0.3212 | 228.5 | 34.8 | 1.02 |
| 50 | our result | 36.57 | 0.3260 | 298.6 | 38.4 | 0.93 |

The results for $\lambda$ approximately follow the expectation based on the geometric model discussed above. The simple geometric model would have given $a_\theta$ =37 g/cm$^2$ and $b_\theta$ =1/3 for collimator angles large enough to accept the elastic scattering. In our case, the nuclear attenuation length also depends on collimator angle. This is due to the angular dependence of the elastic scattering contribution to the removal cross section.

We note that this parameterization gives radiation lengths $X_0$ that are significantly different from the standard values, $a_X$ =716.4 g/cm$^2$ and $b_X$ =287. This may be due to contributions of plural nuclear scattering to the Coulomb multiple scattering distribution, which is not included in the Gaussian approximation used here.

## 5  Summary

The simulation for proton radiography has been implemented with Geant4 toolkit. The simulation results do a good job of reproducing the transmission versus areal density measurements. The inclusion of the nuclear elastic scattering angular distributions is the major piece of physics beyond a simple MCS description of the angular distribution that improves the agreement. In determining volume density distributions by fitting parametric models of the objects to experimental radiographs, the mean free path which depends on the collimator angle and the radiation length are treated as empirical parameters. The mean free path which depends on the collimator angle and the radiation length are treated as empirical parameters.

# 高能质子照相中核弹性散射效应的蒙特卡罗模拟与参数化处理

XU Hai-Bo（许海波）　　ZHENG Na（郑娜）

**摘要**　　基于Geant4程序对质子照相过程进行了模拟，计算分析了核弹性散射对台阶样品的透射率影响，数值模拟和实验结果符合良好。传统的透射率表达式中只考虑核衰减和多次库伦散射，本文根据数值模拟得到的透射率与物质厚度的关系，拟合得到物质的有效平均自由程（与准直角有关）和有效辐射长度表达式，这些结果可用于质子照相的密度重建。

**关键词**：质子照相，库伦散射，角度准直器，核弹性散射，Geant4